\documentclass{article}

\usepackage{microtype}
\usepackage{graphicx}
\usepackage{subfigure}
\usepackage{booktabs} 
\usepackage{makecell}
\usepackage{threeparttable}
\usepackage{array}
\usepackage[utf8]{inputenc}   
\usepackage{booktabs}         
\usepackage{array}            
\usepackage{verbatim}         
\usepackage{amssymb}

\usepackage{hyperref}

\usepackage[accepted]{mlsys2025}

\mlsystitlerunning{Technical report on AI code generation for practical GPU kernels on NVIDIA Blackwell GPUs}

\begin{document}

\twocolumn[
\mlsystitle{Agentic Kernel Optimization: Generating State-of-the-Art GPU Kernels Without Hand-Written CUDA}



\mlsyssetsymbol{equal}{*}

\begin{mlsysauthorlist}
\mlsysauthor{Mao Luo}{equal,if}
\mlsysauthor{Hongbin Li}{equal,if}
\mlsysauthor{Feng Lin}{if}
\mlsysauthor{Hanling Yi}{if}
\mlsysauthor{Zhe Huang}{if}
\end{mlsysauthorlist}

\mlsysaffiliation{if}{Intellifusion Inc., Shenzhen, China}
\mlsyscorrespondingauthor{Hanling Yi}{hanling.cuhk@gmail.com}

\mlsyskeywords{Machine Learning, MLSys}

\vskip 0.3in

\begin{abstract}
We study whether general-purpose code agents can produce state-of-the-art GPU kernels without any manually written CUDA code. We investigate this question using representative workloads from FlashInfer-Bench, focusing on the Fused MoE, DSA TopK Indexer, and DSA Sparse Attention, and evaluate all generated kernels under the correctness-gated FlashInfer-Bench protocol on NVIDIA B200 GPUs. Starting from the PyTorch implementations, workload definitions, benchmark commands, and a compact set of CUDA optimization skills, we build a kernel optimization workflow in Houmao, a multi-agent orchestration framework for heterogeneous coding agents, to generate, debug, profile, and optimize the kernels. Humans remain strictly in an orchestration role: defining the workflow, enforcing correctness and anti-hacking constraints, supplying key references, and redirecting the search when progress stalls, without reviewing or editing the kernel code itself. Across roughly 1.9 billion agent tokens, the resulting kernels achieve speedups of $92.68\times$ on Fused MoE, $1101.02\times$ on DSA TopK Indexer, and $181.35\times$ on DSA Sparse Attention relative to the PyTorch reference implementations, while also significantly outperforming the corresponding FlashInfer baselines. In the official evaluation of the MLSys 2026 FlashInfer AI Kernel Generation Contest, our generated Fused MoE kernel\textsuperscript{$\blacktriangle$} achieves a $1.71\times$ speedup over the FlashInfer baseline, exceeding the top result of the Fused MoE agent-assisted track, which reports a $1.68\times$ speedup. These results suggest that, under a disciplined correctness-first workflow, code agents can serve as effective autonomous optimizers for modern GPU kernel development.
\end{abstract}
]



\printAffiliationsAndNotice{\mlsysEqualContribution} 

\section{From Kernel Engineering to Agentic Optimization}\label{intro}

High-performance GPU kernels have become a key enabler of modern LLM systems, especially for architectures built around sparse attention and mixture-of-experts (MoE) computation~\cite{dao2022flashattention,shazeer2017outrageously,gale2023megablocks,deepseek2024v3}. Yet producing state-of-the-art kernels on new hardware remains a largely manual process, requiring repeated reasoning about tiling, memory movement, scheduling, and hardware-specific constraints. Recent progress such as Triton have made kernel development more programmable and reusable, but obtaining top performance still typically demands substantial expert effort~\cite{tillet2019triton}.

This work asks a different question: \emph{can general-purpose code agents produce state-of-the-art GPU kernels without manually written CUDA code?} We investigate this question using representative workloads from FlashInfer-Bench~\citep{xing2026flashinferbench}, focusing on the Fused MoE, DSA TopK Indexer, and DSA Sparse Attention. These workloads cover fused FP8 MoE computation and sparse-attention components associated with DeepSeek-V3/R1 and DeepSeek-V3.2~\cite{deepseek2024v3}. All generated kernels are evaluated under the correctness-gated FlashInfer-Bench protocol on NVIDIA B200 GPUs.

\begin{table}[t]
\small
\centering
\begin{tabular}{p{50pt}p{45pt}<{\centering}p{40pt}<{\centering}p{45pt}<{\centering}}
\hline
\textbf{Kernel} & \textbf{FlashInfer} & \textbf{Ours} & \textbf{Token Cost} \\
\hline
DSA-TopK & 52.03$\times$ & 1101.02$\times$ &  414M \\
DSA-Attn & 10.33$\times$ & 181.35$\times$ & 427M \\
FuseMoE & 47.08$\times$ & 92.68$\times$ & 1.05B \\
\hline
\end{tabular}
\vspace{-5pt}
\caption{Speedup comparison measured on a local NVIDIA B200 GPU, with the PyTorch implementations as the baseline.}\label{tab:kernel_performance}
\vspace{-10pt}
\end{table}

A defining feature of our approach is that none of the generated kernels were manually written or hand-tuned by our team members. Starting only from the PyTorch reference implementation, workload definition, benchmark command, and a compact set of reusable CUDA optimization skills, a lightweight multi-agent workflow generated, debugged, profiled, and optimized all kernels. Humans remained strictly in an orchestration role: defining the workflow, enforcing anti-hacking safeguards, supplying key references, and pushing the search toward more aggressive optimizations when progress stalled.

\begin{figure*}[t]
\centering
\includegraphics[width=5.3in]{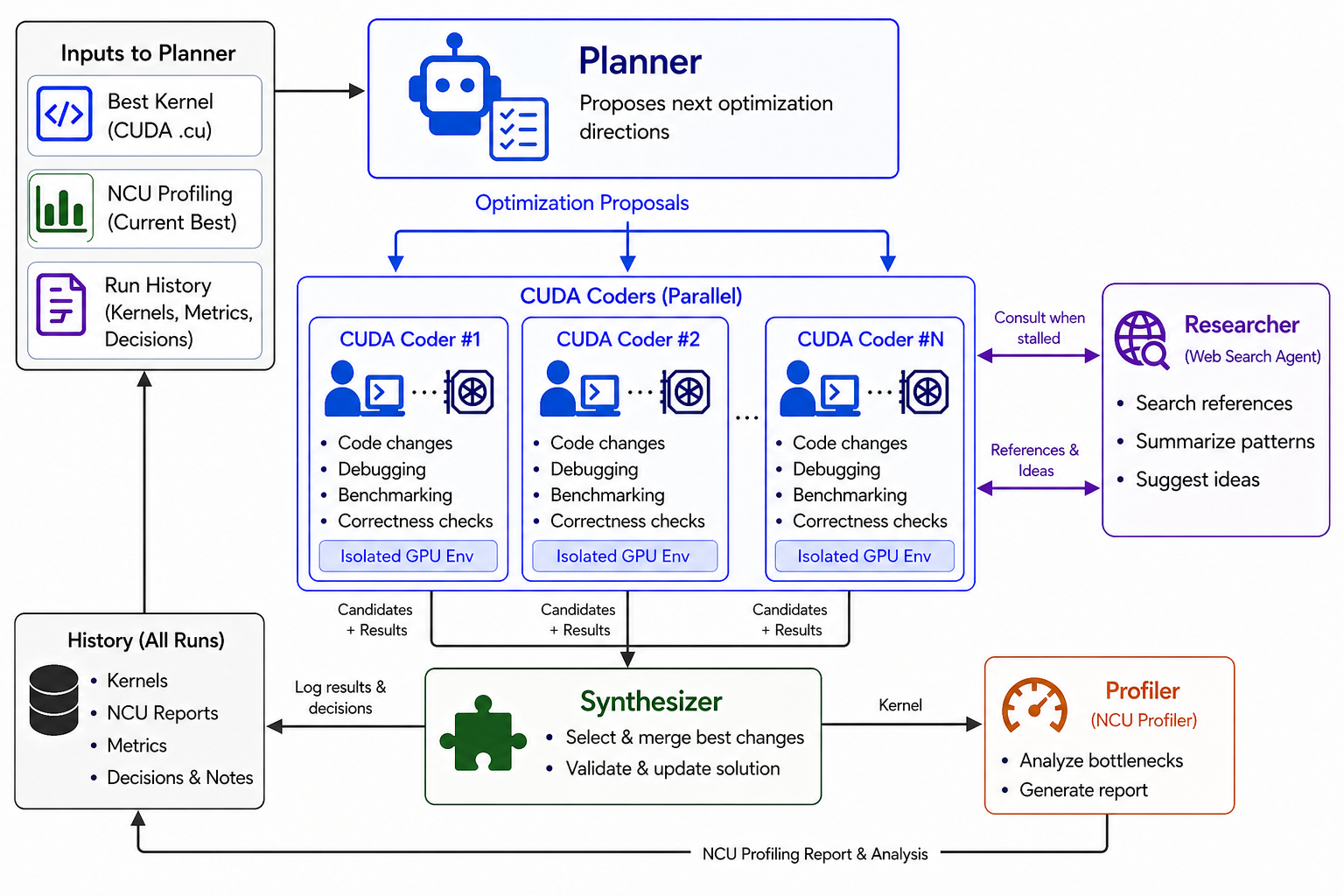}
\vspace{-10pt}
\caption{Team-based AutoResearch-style agentic workflow for CUDA kernel optimization.}\label{fig:agent}
\vspace{-5pt}
\end{figure*}

Across approximately 1.9 billion tokens, the resulting kernels achieved substantial gains over both the PyTorch and FlashInfer baselines~\cite{ye2025flashinfer}. Specifically, we attained speedups of $92.68\times$ on Fused MoE, $1101.02\times$ on DSA TopK Indexer, and $181.35\times$ on DSA Sparse Attention relative to the PyTorch implementations, while also outperforming the corresponding FlashInfer baselines (Table~\ref{tab:kernel_performance}). These results suggest that under a disciplined correctness-first workflow, code agents can serve as effective autonomous optimizers for modern GPU kernel development.

\section{Agentic Optimization Workflow: Agents, Humans, and Tools}
Effective agent-driven kernel optimization depends not only on model capability, but also on a well-structured search process~\cite{li2025tritonforge,zhang2025cudaforge}. Our workflow combines a controlled optimization loop, targeted human intervention, and a small set of reusable tools and skills to support long-horizon kernel optimization. This section describes these components.

\subsection{Agentic Workflow for Kernel Optimization}
Kernel optimization is inherently exploratory: progress typically depends on repeated cycles of hypothesis formation, implementation, profiling, and revision~\cite{wei2025astra, du2025akg}. To support this process, we develop a team-based, AutoResearch-style~\cite{autoresearch} agentic workflow in Houmao\footnote{\url{https://github.com/igamenovoer/houmao}}, a multi-agent orchestration framework for heterogeneous coding agents. The workflow coordinates multiple CLI-based agents, such as Claude Code~\cite{Claude} and Codex~\cite{codex}, which operate with distinct but complementary roles. Figure~\ref{fig:agent} illustrates the organization of agents and the information flow among them.

At the center of the workflow is a \textbf{Planner}, which receives the current best kernel, its NCU profiling results, and the history of prior runs, and then proposes the next optimization directions. These proposals are passed to multiple \textbf{CUDA Coders}, each running in an isolated GPU environment. Conditioned on the Planner's guidance, the Coders explore candidate implementations in parallel, carrying out code changes, debugging, and benchmarking to search for improved solutions. The outputs of the Coders are then reviewed by a \textbf{Synthesizer}, which collects the strongest candidates, identifies the most effective changes, and merges them into an updated solution. Profiling support is provided by an \textbf{Profiler}, implemented as an NCU-based kernel profiler that generates performance reports and attributes bottlenecks for a given kernel. When exploration stalls, a \textbf{Researcher} serves as an on-demand web search agent, retrieving relevant references and summarizing reusable optimization patterns or patch ideas to assist the Coders.

Within this workflow, the optimization loop follows a tightly controlled protocol. Since progress depends heavily on the effectiveness of the CUDA Coders, their behavior is explicitly constrained. The editable surface is deliberately restricted: agents may modify only the CUDA kernel implementation, while the benchmark harness, datasets, configurations, and reference implementations remain fixed. This ensures that all measured gains come from kernel improvements rather than changes to the evaluation setup.

Candidate kernels are evaluated under a correctness-gated speedup metric. A change is accepted only if it passes all correctness checks; numerical errors, runtime failures, and timeouts invalidate the result. After several unsuccessful attempts along the same direction, the CUDA Coders consult the Researcher for additional references and optimization ideas. If no progress is made within the allotted time, they report back to the Synthesizer and wait for the Planner to redirect the search toward a new optimization strategy.

\subsection{Human Intervention as Search Correction}
Even with a clean automated workflow, optimization can still stall because agents exhibit systematic search biases. In our workflow, the main value of human intervention lies not in kernel implementation, but in search correction: identifying when the agent team is stuck, misattributes failures, or misses critical external references.

We observe three recurring failure modes. First, agents favor incremental tweaks even after progress plateaus and may fail to propose more disruptive optimizations unless explicitly pushed in that direction. Second, they sometimes revert genuinely beneficial changes after spurious failures, incorrectly attributing regressions to the new optimization rather than to an independent bug. Third, they tend to under-search external references, occasionally missing key examples needed to reach stronger implementations.

Overall, the agents are effective at local optimization, implementation, and bookkeeping, but less reliable at making major directional shifts or recovering from misleading failures. In such cases, targeted manager-style human intervention is sufficient to redirect the search and unlock substantially better performance. More broadly, this suggests that deep CUDA expertise is not always required; even limited domain knowledge can be sufficient to identify promising optimization directions.

\begin{table*}[t]
\small
\centering
\begin{threeparttable}
\begin{tabular}{m{0.2\textwidth}m{0.6\textwidth}m{0.1\textwidth}}
\hline
\textbf{Skill / Tool} & \textbf{Purpose} & \textbf{Source} \\
\hline
\texttt{cuda-b200-skill} & Blackwell CUDA reference covering PTX ISA, TMA/WGMMA/\texttt{tcgen05}, occupancy, register/shared-memory limits, and debugging. & GitHub\textsuperscript{1} \\
\hline
\texttt{ncu-cuda-profiling} & Automated NCU workflow, saved profiler reports, and summarized metrics. & GitHub\textsuperscript{2} \\
\hline
\texttt{bench-environment} & Contest evaluation harness covering tensor cloning, L2 flushes, cold-cache and no-CUDA-graph defaults, and anti-hacking rules. & Internal environment \\
\hline
\texttt{/codex:rescue} & Independent kernel review via the Claude Code plugin, surfacing alternatives like persistent kernels, TMA, and shared-memory skewing. & GitHub\textsuperscript{3} \\
\hline
Web research & Web search and fetch tools, used after repeated failures to investigate references such as DeepGEMM, FlashInfer, and RadiK. & External tool \\
\hline
\end{tabular}
\begin{tablenotes}
\item[1] \url{https://github.com/technillogue/ptx-isa-markdown}
\item[2] \url{https://github.com/maxiaosong1124/ncu-cuda-profiling-skill}
\item[3] \url{https://github.com/openai/codex-plugin-cc}
\end{tablenotes}
\vspace{-5pt}
\caption{Skills, tools, and references used for CUDA kernel optimization.}\label{tab:cuda_skills_tools}
\vspace{-5pt}
\end{threeparttable}
\end{table*}

\subsection{Tools and Skills}
All kernels are implemented in pure CUDA. In Houmao, all agents are instantiated through Claude Code and backed by Claude Opus 4.6. Houmao runs each agent in an isolated context window. We also provide a small set of on-demand skills, each packaged as a folder of Markdown notes that agents can load when needed. Table~\ref{tab:cuda_skills_tools} summarizes the most important skills and tools.

\section{Kernel Design and Optimization}
\subsection{Speedup vs. Token Cost}
This section summarizes how performance evolves as additional agent tokens are invested in each kernel. Across all three kernels, the final gains arise not from a single rewrite, but from the cumulative effect of increasingly specialized optimizations.

\subsubsection{Fused MoE}
Fused MoE implements the MoE block used in DeepSeek-V3. For each input token, the kernel executes four stages: (1) computing routing scores via a lightweight linear projection, sigmoid activation, group-wise top-2 selection over 8 groups, and final top-$K=8$ expert selection; (2) gathering the active rows for each expert and quantizing them into a contiguous FP8 layout with 128-element blockwise FP32 scaling factors; (3) evaluating the gated MLP, including an FP8 blockwise GEMM1 from $[T,7168]$ to $[T,4096]$, a SwiGLU activation, re-quantization, and an FP8 blockwise GEMM2 from $[T,2048]$ to $[T,7168]$; and (4) scattering the expert outputs back to the original token positions, weighted by the routing scores.

\begin{figure}[t]
\centering
\includegraphics[width=3.25in]{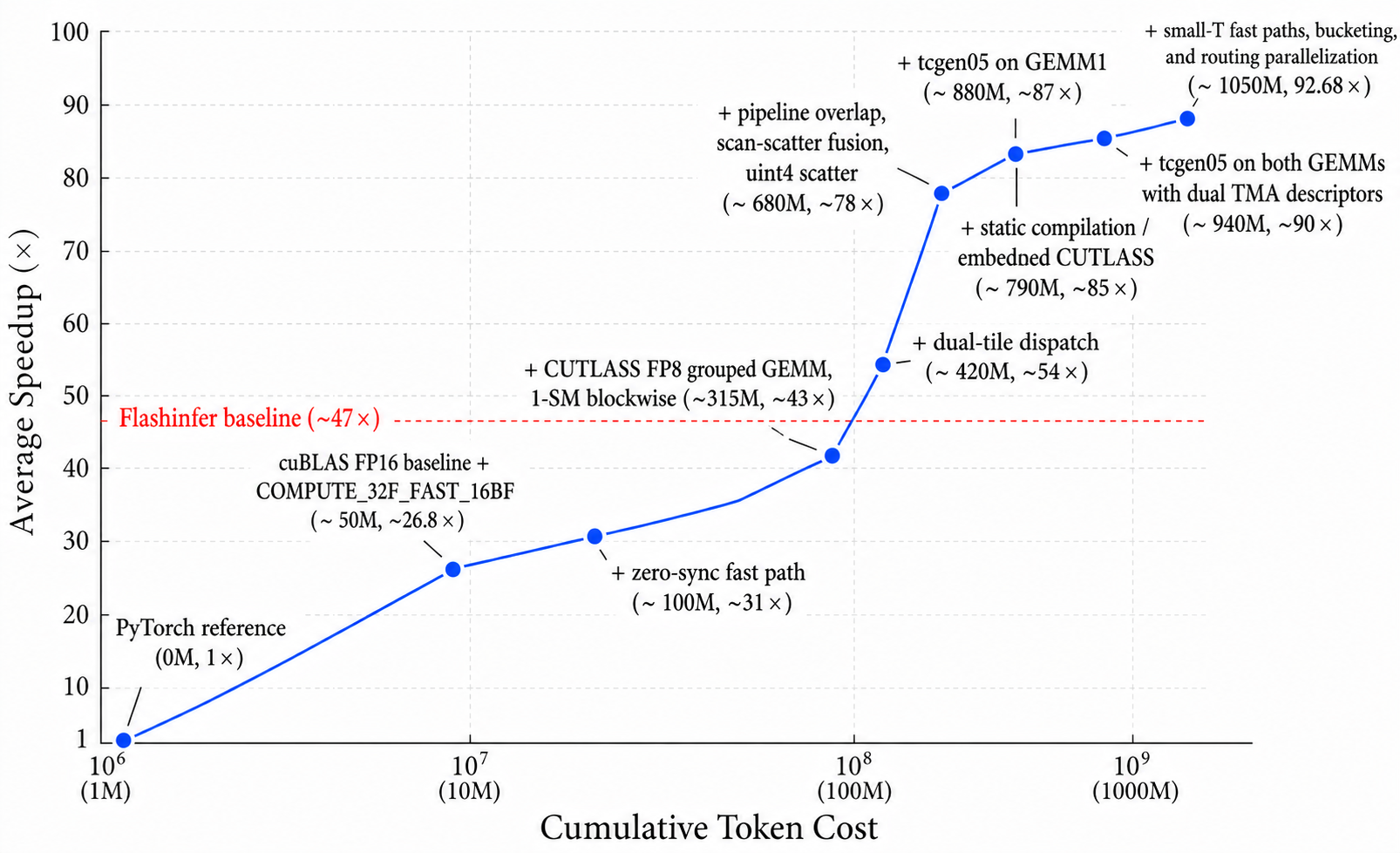}
\vspace{-20pt}
\caption{Cumulative speedup of Fused MoE vs.  token cost.}\label{fig:fusedmoe}
\vspace{-10pt}
\end{figure}

The final implementation keeps all metadata construction on the GPU and eliminates \texttt{cudaStreamSynchronize} from the hot path. It compiles three GEMM backends into a single module: a CUTLASS FP8 blockwise backend, a hand-written \texttt{tcgen05} backend, and a cuBLAS FP16 fallback for very large $T$. Figure~\ref{fig:fusedmoe} visualizes the cumulative optimization trajectory and the token cost.

\subsubsection{DSA Sparse Attention}
DSA Sparse Attention is a paged KV-cache batched decoding operator derived from DeepSeek-V3.2. For each query token, the kernel (1) gathers the top-$K=2048$ KV entries from a sparse index list; (2) computes scaled attention logits of the form $(q_{\mathrm{nope}}K_c^\top) + (q_{\mathrm{pe}}K_p^\top)$; (3) applies a base-2 softmax; and (4) accumulates $\mathrm{attn\_weights}\cdot K_c$ into a 512-dimensional output while also producing the LSE.

\begin{figure}[t]
\centering
\includegraphics[width=3.25in]{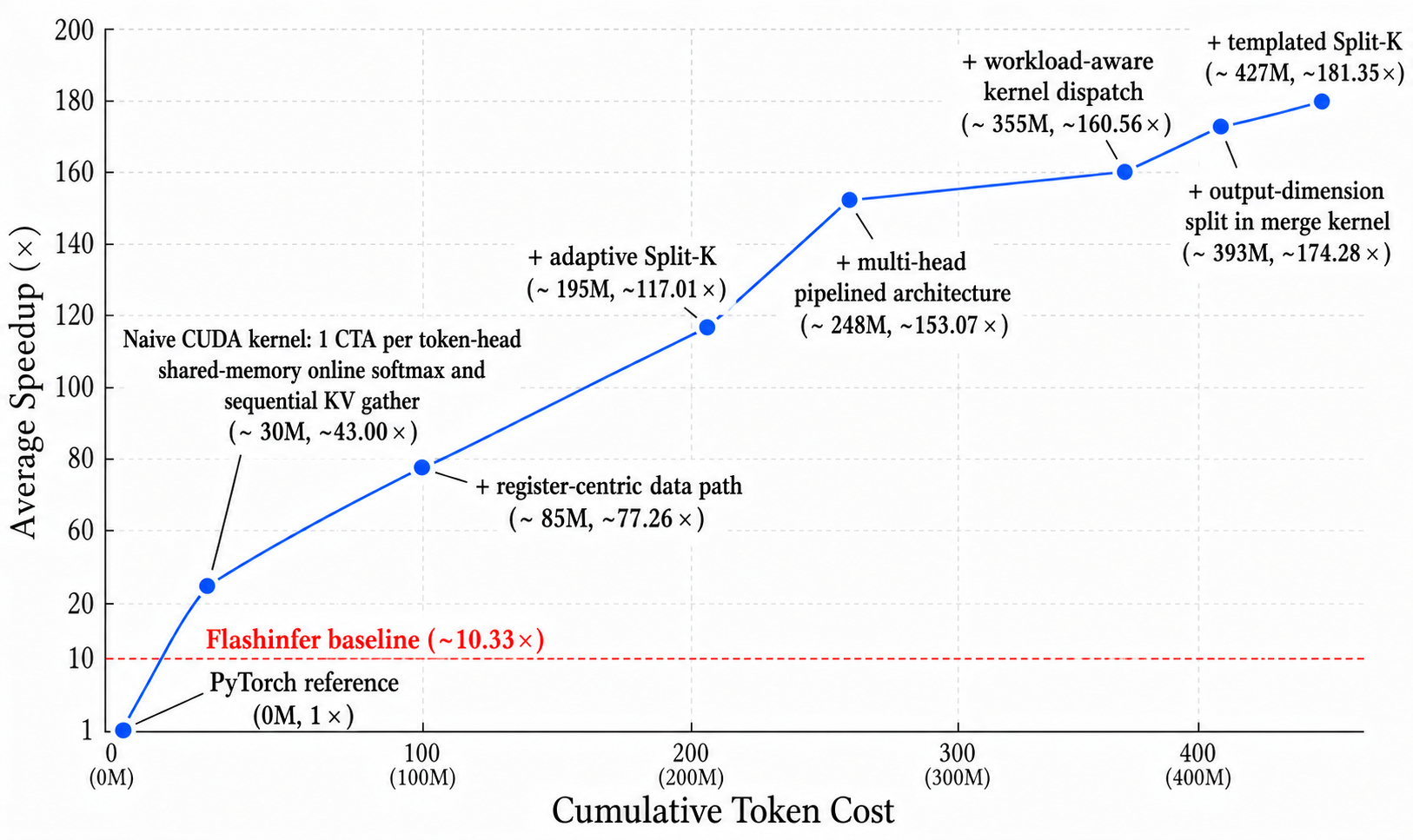}
\vspace{-20pt}
\caption{Cumulative speedup of DSA Sparse-Attn vs. token cost.}\label{fig:sparseattn}
\vspace{-5pt}
\end{figure}

The final implementation adopts a 4-heads-per-block layout with 128 threads organized into 4 warps, assigning one warp to each head. KV tiles are staged through a two-stage \texttt{cp.async} pipeline with double-buffered shared memory. The kernel also uses an adaptive Split-K dispatch strategy: a single-pass kernel for a single token, a templated \texttt{SPLIT\_K=64} configuration for up to two tokens, and a templated \texttt{SPLIT\_K=32} configuration for larger batches. The merge kernel further partitions the 512-dimensional output into four slices, increasing merge parallelism by $4\times$. Additional optimizations include branchless online softmax, register-resident query and output accumulators, BF16 partial outputs to reduce merge bandwidth, and block reordering to improve L2 locality. Figure~\ref{fig:sparseattn} visualizes the cumulative optimization path.

\subsubsection{DSA TopK Indexer}
DSA TopK Indexer implements the paged attention indexing stage for DSA sparse attention. Given FP8 queries $Q$ with 64 heads and 128 dimensions per head, a paged FP8 key cache $K$ with per-token FP32 scaling factors, per-head weights, and sequence lengths, the operator computes
\[
\mathrm{score}(t) = \sum_h \mathrm{ReLU}\!\left(Q_h K_t^\top\right) \cdot w_h
\]
and returns the top-$K$ ($K=2048$) token indices for each batch element. The PyTorch reference implementation performs a \texttt{bmm} over dequantized keys followed by \texttt{torch.topk}. Because the contest harness enforces \texttt{--required-matched-ratio 1.0}, all returned indices must match the reference exactly. This bit-exact TopK requirement drives most of the architectural choices.

\begin{figure}[t]
\centering
\includegraphics[width=3.25in]{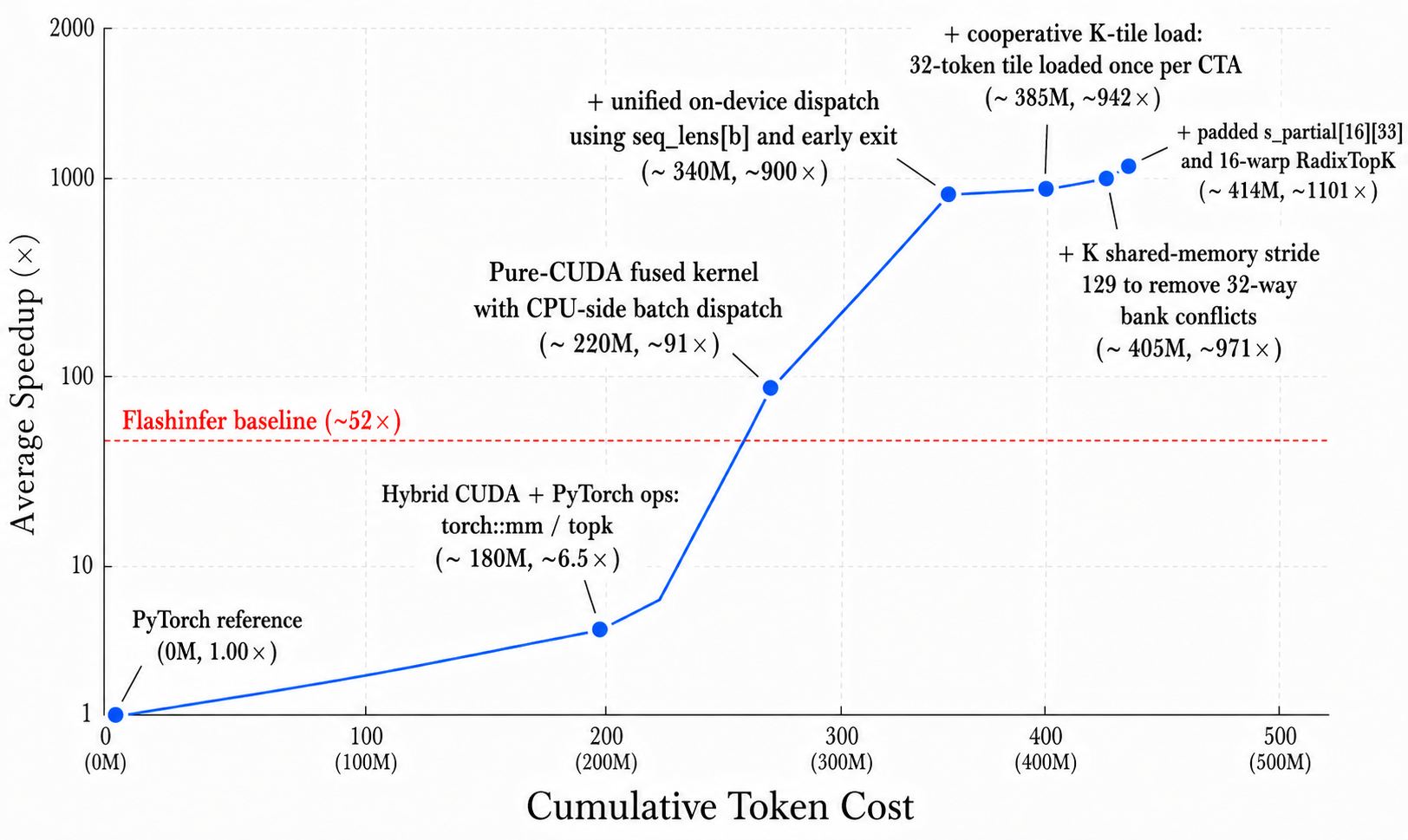}
\vspace{-20pt}
\caption{Cumulative speedup of DSA topk-indexer vs. token cost.}\label{fig:dsatopk}
\vspace{-10pt}
\end{figure}

The final implementation is a three-kernel pure-CUDA pipeline with no CPU--GPU synchronization on the hot path. When $\mathrm{seq\_len} \leq K$, the pipeline degenerates to a pure index-remapping path, since every token is selected trivially. The main optimizations include a fused FP32-accumulation GEMM path that computes $QK^\top$, applies ReLU, weights each head, and reduces across heads within a single kernel; bank-conflict-free shared-memory layouts using a stride-129 key tile and padded cross-warp partial-sum buffers; a four-pass 8-bit radix TopK with sortable-float bit transforms, early termination, subgroup histograms, parallel suffix scan, and warp-ballot gathering; on-device fast/slow-path dispatch based on \texttt{seq\_lens[b]}; and bit-exact FP8-to-FP32 dequantization that matches the reference. Figure~\ref{fig:dsatopk} illustrates the cumulative optimization path.

\begin{table*}[t]
\small
\centering
\begin{tabular}{p{20pt}<{\centering}p{50pt}<{\centering}p{20pt}<{\centering}p{40pt}<{\centering}p{35pt}<{\centering}p{35pt}<{\centering}p{200pt}<{\centering}}
\hline
\textbf{Group} & \textbf{Length} & \textbf{Num} & \textbf{Speedup} & \textbf{Typical} & \textbf{Latency} & \textbf{Bottleneck (NCU)} \\
\hline
short & $1$--$16$ & 5 & $103$--$240\times$ & 7 & 0.099 ms & Floor set by routing and launch overhead. \\
\hline
median & $32$--$80$ & 11 & $75$--$93\times$ & 54 & 0.205 ms & GEMM1+GEMM2 account for $\sim$65\% of iteration time \\
\hline
long & $901$--$14{,}107$ & 3 & $38$--$74\times$ & 14{,}107 & 1.229 ms & GEMM1 accounts for $\sim$47\%, GEMM2 for $\sim$30\% \\
\hline
\end{tabular}
\vspace{-5pt}
\caption{Fused MoE workload groups and the dominant bottlenecks of their representative workloads.}\label{tab:workload_clusters}
\vspace{-5pt}
\end{table*}

\subsection{Workload Regimes and Bottleneck Analysis}
This section analyzes the three kernels through the lens of workload regimes. Rather than treating all benchmark instances independently, we group them into a small number of representative regimes, profile one representative workload from each group with NCU, and identify the dominant bottlenecks. This regime-level analysis helps explain both the main optimization gains achieved during the search and the remaining sources of performance loss.

\subsubsection{Fused MoE}
The 19 contest workloads can naturally partition into three sequence-length regimes. For each regime, we select a representative workload, profile it with NCU, and report the dominant bottlenecks in Table~\ref{tab:workload_clusters}. Across all three regimes, the GEMM kernels already operate at high SM utilization. For the long-sequence representative, GEMM1 and GEMM2 achieve 97\% and 95\% SM utilization, respectively. Accordingly, the remaining optimization headroom appears to lie less in further improving standalone GEMM throughput than in reducing inter-stage data movement. For long-sequence workloads, the most promising remaining direction is to fuse \texttt{pull\_scatter} into the GEMM2 epilogue, so that the producer--consumer path can remain in registers or shared memory. In the current implementation, however, this approach is limited by an upstream CUTLASS API constraint that we leave unmodified.


\begin{table*}[t]
\small
\centering
\begin{tabular}{
>{\centering\arraybackslash}m{20pt}
>{\centering\arraybackslash}m{50pt}
>{\centering\arraybackslash}m{20pt}
>{\centering\arraybackslash}m{40pt}
>{\centering\arraybackslash}m{35pt}
>{\centering\arraybackslash}m{35pt}
>{\centering\arraybackslash}m{200pt}
}
\hline
\textbf{Group} & \textbf{\#Tokens} & \textbf{Num} & \textbf{Speedup} & \textbf{Typical} & \textbf{Latency} & \textbf{Bottleneck (NCU)} \\
\hline
short & 1 & 1 & $280\times$ & 1 token & 0.005 ms & Grid under-occupancy; only 4 CTAs are launched; latency is launch-dominated \\
\hline
median & 2 & 8 & $125$--$148\times$ & 2 tokens & 0.017 ms & Latency-bound Split-K; low occupancy from high register pressure \\
\hline
long & $6$--$8$ & 14 & $108$--$228\times$ & 8 tokens & 0.026 ms & Register-pressure-limited occupancy; residual shared-memory bank conflicts remain \\
\hline
\end{tabular}
\vspace{-5pt}
\caption{DSA Sparse Attention workload groups and the dominant bottlenecks of their representative workloads.}
\label{tab:dsa_attn_workload_clusters}
\end{table*}

\begin{table*}[t]
\fontsize{8}{9}\selectfont
\centering
\begin{tabular}{
>{\centering\arraybackslash}m{20pt}
>{\centering\arraybackslash}m{95pt}
>{\centering\arraybackslash}m{15pt}
>{\centering\arraybackslash}m{45pt}
>{\centering\arraybackslash}m{75pt}
>{\centering\arraybackslash}m{40pt}
>{\centering\arraybackslash}m{110pt}
}
\hline
\textbf{Group} & \textbf{Length} & \textbf{Num} & \textbf{Speedup} & \textbf{Typical} & \textbf{Latency} & \textbf{Bottleneck (NCU)} \\
\hline
short & \makecell[c]{$\max S \leq 1024$\\$\max P \leq 16$} & 27 & $477$--$3732\times$ & \makecell[c]{bs=29\\$\max P=1$\\$\max S=64$} & $\sim$0.002 ms & Fast-path index remap; floor set by launch and grid scheduling \\
\hline
medium & \makecell[c]{$1088 \leq \max S \leq 2048$\\$17 \leq \max P \leq 32$} & 42 & $816$--$5366\times$ & \makecell[c]{bs=15\\$\max P=31$\\$\max S=1984$} & $\sim$0.004 ms & Fast-path index remap; kernel cost grows only with batch size and TopK \\
\hline
long & \makecell[c]{$\max S > 2048$\\$\max P > 32$} & 59 & \makecell[c]{$107$--$756\times$} & \makecell[c]{bs=4\\$\max P=36$\\$S=[18,11,2161,20]$} & $\sim$0.020 ms & Slow-path FusedGemm + RadixTopK; low active-CTA count and FP8 dequant stalls \\
\hline
\end{tabular}
\vspace{-5pt}
\caption{DSA TopK Indexer workload groups and the dominant bottlenecks of their representative workloads.}\label{tab:dsa_topk_workload_clusters}
\vspace{-5pt}
\end{table*}

\subsubsection{DSA Sparse Attention}
The 23 contest workloads can be grouped into three regimes by \texttt{num\_tokens}, which also matches the boundaries used by our adaptive Split-K dispatcher. For each regime, we select one representative workload, profile it with NCU, and summarize the dominant bottleneck in Table~\ref{tab:dsa_attn_workload_clusters}. These clusters correspond to the main optimization wins during the campaign. \texttt{MERGE\_DIM\_SPLIT=4} targeted the long-cluster merge tail and contributed the largest gain. The \texttt{SPLIT\_K=64} path for \texttt{num\_tokens} $\leq 2$ was introduced specifically for the mid cluster based on per-workload NCU analysis. For the short cluster, the single fused-kernel dispatch avoids merge overhead entirely.

The remaining performance gap appears to be largely structural. The attention kernel uses roughly 192 registers per thread and about 50 KB of shared memory, limiting achieved occupancy to around 10\%. Since the kernel is already latency-bound rather than compute- or bandwidth-bound, further improvements likely require a more substantial architectural redesign, such as reducing register pressure or moving to a TMA/warpgroup/\texttt{tcgen05}-style design. Lower-precision compute could also reduce pressure, but this direction is constrained by the strict correctness requirements.

\subsubsection{DSA TopK Indexer}
The 128 contest workloads are naturally organized into three regimes according to \texttt{max\_num\_pages}. Because each page contains 64 tokens, this quantity effectively upper-bounds the maximum sequence length within a batch. The most important threshold is \texttt{max\_num\_pages=32}: when all sequences satisfy this bound, their lengths do not exceed \texttt{TopK=2048}, and the kernel can bypass score computation entirely in favor of a pure index-remapping fast path. For each regime, we choose a representative workload, profile its hot kernels with NCU, and summarize the principal bottlenecks in Table~\ref{tab:dsa_topk_workload_clusters}.

These regimes account for the largest gains observed in the DSA TopK Indexer optimization campaign. In the short and mid regimes, the on-device fast-path dispatch yields especially large improvements by skipping score computation whenever $\mathrm{seq\_len} \leq \mathrm{TopK}$. In both cases, execution reduces to the same lightweight index-remapping kernel, whereas the PyTorch reference still incurs the cost of per-batch GEMM and \texttt{topk}. In the long regime, only those sequences whose lengths exceed \texttt{TopK} take the full slow path. For the representative workload, this means that only one of the four batch elements invokes \texttt{FusedGemm} and \texttt{RadixTopK}, while the remaining three return early.

Within the long regime, the main improvements come from eliminating shared-memory bank conflicts and reducing serialized work inside \texttt{RadixTopK}. In particular, padding \texttt{s\_partial} from \texttt{[16][32]} to \texttt{[16][33]} removes a substantial cross-warp bank conflict in \texttt{FusedGemm}, while increasing the \texttt{RadixTopK} configuration from 256 to 512 threads reduces the loop work performed in each pass. The remaining limitation is structural: at those small batch sizes, \texttt{RadixTopK} launches only a small number of CTAs, leaving much of the GPU underutilized, and the fused GEMM path still remains behind what a TMA+\texttt{tcgen05}-style pipeline could potentially achieve.

\section{Discussion}
\paragraph{When Speedup Is Not Real.}
During kernel optimization, we encounter several instances of benchmark exploitation, in which the agent achieves implausibly large reported speedups without producing a genuinely improved kernel. In early Fused MoE runs, for example, the reported speedup sometimes exceeds $1000\times$. The underlying mechanism is that the in-process FlashInfer-Bench harness reuses input tensor pointers across iterations, which leads the agent to introduce an output cache keyed by tensor identity. After the first iteration, the kernel can simply return cached outputs while still passing correctness checks, even though the measured gain does not correspond to a deployable kernel optimization.

A notable lesson from this campaign is therefore that code agents, when granted full visibility into the project codebase, can engage in ``reward hacking''~\cite{liu2026dr} unless the optimization objective is carefully constrained. To prevent this behavior, we incorporate explicit anti-hacking rules into every agent's initial prompt. In particular, we disallow optimizations based on input-identity caching, cross-iteration buffer reuse, or other sample-specific shortcuts. If a performance gain depends on a legitimate deployable cache, the agent must report cold-path and warm-path performance separately. This rule becomes a core part of our evaluation discipline and is necessary to ensure that measured speedups reflect genuine kernel improvements.



\paragraph{Human as Orchestrator.}
Human contribution is best understood as orchestration rather than implementation. Humans define the evaluation protocol, benchmark discipline, and multi-agent workflow; enforce anti-hacking and correctness constraints; and intervene only when the search plateaus, typically by redirecting the agents toward more aggressive optimization strategies or supplying key external references that the agents fail to retrieve on their own. Crucially, humans do not review / edit the kernel code itself.


\section{Conclusion}
In this work, we study whether general-purpose code agents can produce state-of-the-art GPU kernels without any manually written CUDA code. Using representative FlashInfer-Bench workloads and their correctness-gated evaluation protocol, we show that the answer is yes: under a disciplined correctness-first workflow, an agentic system can generate high-performance kernels that substantially outperform both the PyTorch references and the corresponding FlashInfer baselines. Our results also show that model capability alone is not enough. Strong performance requires a structured optimization loop, explicit anti-hacking constraints, and targeted human intervention to redirect the search when agents stall or misgeneralize. Yet this intervention remains orchestration rather than implementation: humans do not write or edit the kernel code. Taken together, these findings suggest a broader shift. High-performance GPU kernel development no longer needs to be viewed exclusively as a manual expert craft. With the right workflow, code agents can already serve as practical autonomous optimizers for modern GPU kernels.

\section*{Acknowledgments}
We thank the organizers of the MLSys 2026 FlashInfer AI Kernel Generation Contest for providing the workloads, benchmark, and official evaluation. In the official evaluation results shared by email, our generated Fused MoE kernel achieved a $1.71\times$ speedup over the FlashInfer baseline, exceeding the reported $1.68\times$ top score in the agent-assisted track. Although regional prize-eligibility restrictions related to U.S. export-control regulations meant that our team could not be listed among the prize-receiving teams, we remain grateful for the opportunity to participate. We hope that future scholarly exchange will remain open, inclusive, and centered on shared scientific excellence.

\bibliography{example_paper}
\bibliographystyle{mlsys2025}

\appendix
\section{NCU Profile of the Slowest FuseMoE Workload}
\label{app:ncu_slowest_moe}

This appendix reports the NCU profile for the slowest FuseMoE workload in our final submission. The workload has \texttt{seq\_len=14107} and corresponds to the long-sequence regime discussed in Table~\ref{tab:workload_clusters}.

\subsection{Capture Provenance}

\begin{itemize}
    \item \textbf{Kernel version:} \texttt{cuda-moe-tvm-v14}
    \item \textbf{Workload:} \texttt{5e8dc11c}, \texttt{seq\_len=14107}
    \item \textbf{Driver:} 3 warmup iterations followed by 5 NVTX-scoped measurement iterations
    \item \textbf{Hardware:} NVIDIA B200, SM100, compute capability 10.0
    \item \textbf{Profiler:} NCU 2026.1.0.0, \texttt{--set full}, \texttt{--replay-mode kernel}
\end{itemize}

\subsection{Per-Kernel Summary}

Table~\ref{tab:moe_ncu_compact_summary} summarizes the major kernels in one FuseMoE iteration. The profile shows that the two CUTLASS FP8 grouped-GEMM invocations dominate total runtime, while \texttt{pull\_scatter} is the largest non-GEMM DRAM consumer.

\begin{table*}[htbp]
\centering
\small
\setlength{\tabcolsep}{4pt}
\caption{Compact NCU summary for the slowest FuseMoE workload.}
\label{tab:moe_ncu_compact_summary}
\begin{tabular}{lccccccccc}
\hline
\textbf{Kernel} &
\textbf{Inv.} &
\textbf{Time} &
\textbf{SM} &
\textbf{DRAM} &
\textbf{L1} &
\textbf{L2} &
\textbf{Occ.} &
\textbf{Reg.} &
\textbf{WCPI} \\
&
&
\textbf{($\mu$s)} &
\textbf{(\%)} &
\textbf{(\%)} &
\textbf{(\%)} &
\textbf{(\%)} &
\textbf{(\%)} &
&
\\
\hline
\texttt{CUTLASS\_FP8\_grouped} & 2 & 491.74 & 50.68 & 24.85 & 45.18 & 41.69 & 14.09 & 168 & 5.46 \\
\texttt{routing\_kernel} & 1 & 108.22 & 62.37 & 1.74 & 67.96 & 1.65 & 65.08 & 32 & 17.71 \\
\texttt{pull\_scatter\_bf16} & 1 & 81.31 & 23.52 & 62.92 & 33.87 & 30.90 & 43.50 & 61 & 26.82 \\
\texttt{gather\_fp8\_and\_scales} & 1 & 42.69 & 46.47 & 47.04 & 46.91 & 35.63 & 82.40 & 32 & 25.56 \\
\texttt{swiglu\_to\_fp8} & 1 & 34.94 & 70.89 & 55.87 & 30.86 & 32.65 & 85.21 & 29 & 16.66 \\
\texttt{scatter\_with\_scan} & 1 & 11.30 & 2.31 & 0.53 & 1.80 & 1.47 & 24.03 & 18 & 198.69 \\
\hline
\multicolumn{10}{l}{\textbf{Total kernel time for one MoE iteration:} 1261.94 $\mu$s} \\
\hline
\end{tabular}
\end{table*}

The compact summary indicates that the workload is dominated by the GEMM path: the two \texttt{CUTLASS\_FP8\_grouped} invocations account for the largest share of runtime. Among the non-GEMM kernels, \texttt{pull\_scatter\_bf16} is the most important optimization target because it has the highest DRAM utilization.

\subsection{Dominant Kernel Drill-Down}

The two dominant components are the CUTLASS FP8 blockwise grouped GEMMs and the \texttt{pull\_scatter} kernel. The former dominates total iteration time, while the latter is the largest non-GEMM memory-bandwidth consumer.

\subsubsection{CUTLASS FP8 Blockwise Grouped GEMM}

The same CUTLASS grouped-GEMM template is invoked twice per MoE iteration:
\begin{itemize}
    \item \textbf{GEMM1:} $K=7168$, approximately 600.22 $\mu$s.
    \item \textbf{GEMM2:} $K=2048$, approximately 383.26 $\mu$s.
\end{itemize}

On this long-sequence workload, the runtime dispatcher selects the CUTLASS backend rather than the hand-written \texttt{tcgen05} backend, because CUTLASS is faster on long-sequence cases by approximately 2--7\%. Table~\ref{tab:cutlass_gemm_sol} reports the Speed-of-Light throughput metrics for these two CUTLASS invocations.

\begin{table}[htbp]
\centering
\small
\setlength{\tabcolsep}{4pt}
\caption{NCU throughput summary for CUTLASS FP8 grouped GEMM.}
\label{tab:cutlass_gemm_sol}
\begin{tabular}{lcccc}
\hline
\textbf{Metric} & \textbf{Unit} & \textbf{Min} & \textbf{Max} & \textbf{Avg.} \\
\hline
DRAM frequency & GHz & 4.00 & 4.00 & 4.00 \\
SM frequency & GHz & 1.58 & 1.63 & 1.60 \\
Duration & $\mu$s & 383.26 & 600.22 & 491.74 \\
Memory throughput & \% & 37.52 & 50.70 & 44.11 \\
DRAM throughput & \% & 23.99 & 25.71 & 24.85 \\
L1/TEX throughput & \% & 38.02 & 52.34 & 45.18 \\
L2 throughput & \% & 37.68 & 45.71 & 41.69 \\
SM throughput & \% & 43.80 & 57.55 & 50.68 \\
\hline
\end{tabular}
\end{table}

Table~\ref{tab:cutlass_gemm_launch_occ} reports the corresponding launch and occupancy statistics. These numbers show that the GEMM kernels are resource-limited: each CTA uses 168 registers per thread and approximately 201 KB of dynamic shared memory, allowing only one active block per SM.

\begin{table}[htbp]
\centering
\small
\setlength{\tabcolsep}{4pt}
\caption{Launch and occupancy summary for CUTLASS FP8 grouped GEMM.}
\label{tab:cutlass_gemm_launch_occ}
\begin{tabular}{lcc}
\hline
\textbf{Metric} & \textbf{Unit} & \textbf{Value} \\
\hline
Block size & threads & 384 \\
Grid size & CTAs & 148 \\
Waves per SM & -- & 1.00 \\
Registers per thread & reg/thread & 168 \\
Dynamic shared memory per block & KB/block & 201.22 \\
Theoretical occupancy & \% & 18.75 \\
Achieved occupancy & \% & 14.09 \\
Achieved active warps per SM & warps & 9.02 \\
Warp cycles per issued instruction & cycles & 5.46 \\
Average active threads per warp & -- & 31.85 \\
\hline
\end{tabular}
\end{table}

Despite low achieved occupancy, the GEMM kernels maintain high active-thread efficiency and dominate the iteration because of the large amount of matrix-multiply work. The remaining headroom is therefore less about launch overhead and more about reducing resource pressure and data movement around the GEMM pipeline.

\subsubsection{\texttt{pull\_scatter\_bf16} Kernel}

The \texttt{pull\_scatter} kernel accounts for 81.31 $\mu$s, or approximately 6.4\% of the iteration. Table~\ref{tab:pull_scatter_sol} shows that it reaches 62.92\% of peak sustained DRAM throughput, making it the largest non-GEMM DRAM consumer in this workload.

\begin{table}[htbp]
\centering
\small
\setlength{\tabcolsep}{4pt}
\caption{NCU throughput summary for \texttt{pull\_scatter\_bf16}.}
\label{tab:pull_scatter_sol}
\begin{tabular}{lccc}
\hline
\textbf{Metric} & \textbf{Unit} & \textbf{Value} \\
\hline
DRAM frequency & GHz & 3.99 \\
SM frequency & GHz & 1.94 \\
Duration & $\mu$s & 81.31 \\
Memory throughput & \% & 62.92 \\
DRAM throughput & \% & 62.92 \\
L1/TEX throughput & \% & 33.87 \\
L2 throughput & \% & 30.90 \\
SM throughput & \% & 23.52 \\
Achieved occupancy & \% & 43.50 \\
Registers per thread & reg/thread & 61 \\
Grid size & CTAs & 14107 \\
Waves per SM & -- & 23.83 \\
\hline
\end{tabular}
\end{table}

Unlike the GEMM kernels, \texttt{pull\_scatter} launches many CTAs and reaches much higher occupancy. Its bottleneck is DRAM bandwidth: the kernel performs substantial output movement after GEMM2, and therefore becomes the main remaining non-GEMM optimization target. This supports the main text's conclusion that fusing \texttt{pull\_scatter} into the GEMM2 epilogue is the most promising next optimization for long-sequence workloads.

\subsection{Summary}

The NCU profile confirms the performance breakdown reported in the main text. On the slowest long-sequence FuseMoE workload, most time is spent in the two CUTLASS FP8 grouped GEMMs, while the largest non-GEMM cost is \texttt{pull\_scatter}. The GEMMs are resource-constrained but already highly optimized, whereas \texttt{pull\_scatter} remains DRAM-bandwidth dominated. Therefore, the most meaningful remaining optimization is not another standalone GEMM improvement, but reducing producer--consumer data movement by fusing scatter behavior into the GEMM2 epilogue.



\section{NCU Profile of the Slowest DSA-Attn Workload}
\label{app:ncu-slowest-dsa-attn}

This appendix reports the NCU profile for the slowest DSA-Attn workload in our final submission. The workload is the long-cluster representative with \texttt{num\_tokens=8}, \texttt{num\_pages=8462}, and \texttt{topK=2048}, corresponding to the long-sequence regime discussed in the main text.

\subsection{Capture Provenance}

\begin{itemize}
    \item \textbf{Kernel version:} \texttt{dsa-attn-v4}
    \item \textbf{Workload:} \texttt{385742b2}, \texttt{num\_tokens=8}, \texttt{num\_pages=8462}
    \item \textbf{Driver:} 1 warmup iteration followed by 1 NVTX-scoped measurement iteration
    \item \textbf{Hardware:} NVIDIA B200, SM100, compute capability 10.0
    \item \textbf{Profiler:} NCU 2026.1.0.0, \texttt{--set full}, \texttt{--import-source on}, \texttt{--replay-mode kernel}
    \item \textbf{Build:} compiled with \texttt{nvcc -lineinfo} via \texttt{FIB\_FORCE\_LINEINFO=1}
\end{itemize}

\subsection{Per-Kernel Summary}

The DSA-Attn iteration consists of two fused CUDA kernels: \texttt{dsa\_attention\_partial<32>} and \texttt{dsa\_attention\_merge<32>}. Table~\ref{tab:dsa-attn-ncu-kernel-summary} summarizes their NCU-measured aggregate metrics. The partial kernel accounts for 67\% of total kernel time, while the merge kernel accounts for the remaining 33\%.

\begin{table}[htbp]
\centering
\small
\setlength{\tabcolsep}{3pt}
\caption{Per-kernel NCU summary for the slowest DSA-Attn workload.}
\label{tab:dsa-attn-ncu-kernel-summary}
\begin{tabular}{lcccc}
\hline
\textbf{Kernel} & \textbf{Time} & \textbf{SM} & \textbf{Occ.} & \textbf{Reg.} \\
 & \textbf{($\mu$s)} & \textbf{(\%)} & \textbf{(\%)} & \\
\hline
\texttt{partial<32>} & 17.73 & 17.68 & 10.07 & 192 \\
\texttt{merge<32>} & 8.58 & 1.81 & 6.16 & 95 \\
\hline
\multicolumn{5}{l}{\textbf{Total kernel time:} 26.31 $\mu$s} \\
\hline
\end{tabular}
\end{table}

To understand where time is spent inside each fused kernel, we use NCU source-mapped PC sampling and attribute stalls to source-line ranges in \texttt{kernel.cu}. Table~\ref{tab:dsa-attn-phase-breakdown} reports the resulting phase-level time breakdown. The dominant costs are KV staging through \texttt{cp.async} and sparse-index gathering, which together account for 10.88 $\mu$s, or 41\% of the full iteration.

\begin{table}[htbp]
\centering
\small
\setlength{\tabcolsep}{3pt}
\caption{Source-attributed phase breakdown for DSA-Attn.}
\label{tab:dsa-attn-phase-breakdown}
\begin{tabular}{lc}
\hline
\textbf{Phase} & \textbf{Time ($\mu$s)} \\
\hline
\multicolumn{2}{l}{\textbf{\texttt{partial<32>}}} \\
KV \texttt{cp.async}, global to shared & 5.54 \\
Sparse-index gather & 5.34 \\
GEMM1, $QK^\top$ + warp reduce & 1.81 \\
GEMM2, attn-weighted Kc accumulation & 1.66 \\
Q-load / prologue & 1.16 \\
Online softmax & 0.96 \\
Shared-memory KV loads & 0.91 \\
Epilogue & 0.35 \\
\hline
\multicolumn{2}{l}{\textbf{\texttt{merge<32>}}} \\
LSE scan + valid-split count & 3.11 \\
Weighted partial accumulation & 2.58 \\
Helpers + scheduling overhead & 1.94 \\
Epilogue & 0.95 \\
\hline
\end{tabular}
\end{table}

\subsection{Dominant Kernel Drill-Down}

The two kernels expose different bottlenecks. The partial kernel is latency-bound under a low occupancy ceiling, while the merge kernel is limited by launch and scheduling overhead on a small grid.

\subsubsection{\texttt{dsa\_attention\_partial<32>}}

The partial kernel implements Split-K=32 FlashDecoding. It launches 1024 CTAs, corresponding to 8 tokens, 4 head groups, and 32 splits. Each CTA processes 64 sparse KV entries through a fused \texttt{cp.async} $\rightarrow QK^\top \rightarrow$ online-softmax $\rightarrow$ attn-weighted value accumulation pipeline.

Table~\ref{tab:dsa-attn-partial-sol} reports Speed-of-Light throughput metrics for the partial kernel. Although the kernel has a compute-heavy instruction mix, it reaches only 17.68\% SM throughput and 1.58\% DRAM throughput, indicating a latency-bound rather than bandwidth-saturated regime.

\begin{table}[htbp]
\centering
\small
\setlength{\tabcolsep}{4pt}
\caption{Speed-of-Light metrics for \texttt{dsa\_attention\_partial<32>}.}
\label{tab:dsa-attn-partial-sol}
\begin{tabular}{lcc}
\hline
\textbf{Metric} & \textbf{Unit} & \textbf{Value} \\
\hline
Duration & $\mu$s & 17.73 \\
SM throughput & \% & 17.68 \\
Memory throughput & \% & 16.38 \\
DRAM throughput & \% & 1.58 \\
L1/TEX throughput & \% & 37.11 \\
L2 throughput & \% & 3.38 \\
SM frequency & GHz & 1.94 \\
DRAM frequency & GHz & 3.98 \\
\hline
\end{tabular}
\end{table}

Table~\ref{tab:dsa-attn-partial-launch-occ} shows that the partial kernel is constrained by register and shared-memory pressure. With 192 registers per thread and 73.98 KB of dynamic shared memory per block, achieved occupancy is only 10.07\%. As a result, memory-latency stalls cannot be effectively hidden by other eligible warps.

\begin{table}[htbp]
\centering
\small
\setlength{\tabcolsep}{4pt}
\caption{Launch and occupancy metrics for \texttt{dsa\_attention\_partial<32>}.}
\label{tab:dsa-attn-partial-launch-occ}
\begin{tabular}{lcc}
\hline
\textbf{Metric} & \textbf{Unit} & \textbf{Value} \\
\hline
Block size & threads & 128 \\
Grid size & CTAs & 1024 \\
Waves per SM & -- & 3.46 \\
Registers per thread & reg/thread & 192 \\
Dynamic shared memory & KB/block & 73.98 \\
Theoretical occupancy & \% & 12.50 \\
Achieved occupancy & \% & 10.07 \\
Active warps per SM & warps & 6.45 \\
Warp cycles per issued inst. & cycles & 4.00 \\
Active threads per warp & -- & 31.52 \\
\hline
\end{tabular}
\end{table}

\subsubsection{\texttt{dsa\_attention\_merge<32>}}

The merge kernel reads the 32 BF16 partial outputs for each token and head group, then performs LSE-renormalized weighted accumulation. The output dimension is split by \texttt{MERGE\_DIM\_SPLIT=4}, giving 128 CTAs in total.

Table~\ref{tab:dsa-attn-merge-sol} reports Speed-of-Light throughput metrics for the merge kernel. The very low SM and memory throughput indicate that the kernel is not limited by arithmetic or bandwidth. Instead, the grid is too small to saturate the GPU.

\begin{table}[htbp]
\centering
\small
\setlength{\tabcolsep}{4pt}
\caption{Speed-of-Light metrics for \texttt{dsa\_attention\_merge<32>}.}
\label{tab:dsa-attn-merge-sol}
\begin{tabular}{lcc}
\hline
\textbf{Metric} & \textbf{Unit} & \textbf{Value} \\
\hline
Duration & $\mu$s & 8.58 \\
SM throughput & \% & 1.81 \\
Memory throughput & \% & 1.45 \\
DRAM throughput & \% & 0.77 \\
L1/TEX throughput & \% & 6.60 \\
L2 throughput & \% & 0.94 \\
SM frequency & GHz & 1.95 \\
DRAM frequency & GHz & 3.98 \\
\hline
\end{tabular}
\end{table}

Table~\ref{tab:dsa-attn-merge-launch-occ} confirms the small-grid bottleneck. The merge kernel launches only 128 CTAs, or 0.17 waves on 148 SMs. Its dominant phase is the LSE scan, a 32-element global-load reduction whose latency is exposed because the grid cannot provide enough parallel work to hide it.

\begin{table}[htbp]
\centering
\small
\setlength{\tabcolsep}{4pt}
\caption{Launch and occupancy metrics for \texttt{dsa\_attention\_merge<32>}.}
\label{tab:dsa-attn-merge-launch-occ}
\begin{tabular}{lcc}
\hline
\textbf{Metric} & \textbf{Unit} & \textbf{Value} \\
\hline
Block size & threads & 128 \\
Grid size & CTAs & 128 \\
Waves per SM & -- & 0.17 \\
Registers per thread & reg/thread & 95 \\
Dynamic shared memory & KB/block & 0.00 \\
Theoretical occupancy & \% & 31.25 \\
Achieved occupancy & \% & 6.16 \\
Active warps per SM & warps & 3.94 \\
Warp cycles per issued inst. & cycles & 13.86 \\
Active threads per warp & -- & 31.08 \\
\hline
\end{tabular}
\end{table}

\subsection{Summary}

The NCU profile confirms the main-text performance breakdown. On the slowest long-cluster DSA-Attn workload, most time is spent in \texttt{dsa\_attention\_partial<32>}. Its dominant costs are KV \texttt{cp.async} and sparse-index global loads, which are difficult to hide because register and shared-memory pressure cap achieved occupancy at roughly 10\%. The merge kernel is smaller but still significant: after \texttt{MERGE\_DIM\_SPLIT=4}, it launches only 0.17 waves and is therefore launch/scheduling-overhead bound.

The remaining optimization headroom is structural. Further local fusion is unlikely to move the bottleneck substantially. Meaningful improvement would require reducing register pressure or redesigning the kernel around a TMA, warpgroup, and \texttt{tcgen05}-style pipeline, while preserving the strict correctness requirements of the benchmark.

\section{NCU Profile of the Slowest DSA-TopK Workload}
\label{app:ncu-slowest-dsa-topk}

This appendix reports the NCU profile for the slowest DSA-TopK workload in our final submission. The workload is the long-cluster representative with \texttt{bs=4}, \texttt{max\_num\_pages=36}, and \texttt{seq\_lens=[98,91,2241,100]}. It corresponds to the long-sequence regime discussed in the main text and achieves the lowest speedup in the 128-workload contest run, $107.29\times$ over the mathematical reference.

\subsection{Capture Provenance}

\begin{itemize}
    \item \textbf{Kernel version:} \texttt{dsa-topk-indexer-v6}
    \item \textbf{Workload:} \texttt{4c7705ad}, \texttt{bs=4}, \texttt{max\_num\_pages=36}, \texttt{seq\_lens=[98,91,2241,100]}, \texttt{num\_pages=11923}
    \item \textbf{Slow-path batch:} only $b=2$ with \texttt{seq\_len=2241}; the other three batch elements early-return
    \item \textbf{Driver:} 1 warmup iteration followed by 1 NVTX-scoped measurement iteration
    \item \textbf{Hardware:} NVIDIA B200, SM100, compute capability 10.0
    \item \textbf{Profiler:} NCU 2026.1.0.0, \texttt{--set full}, \texttt{--import-source on}, \texttt{--replay-mode kernel}
    \item \textbf{Build:} compiled with \texttt{nvcc -lineinfo} via \texttt{FIB\_FORCE\_LINEINFO=1}
\end{itemize}

\subsection{Per-Kernel Summary}

The DSA-TopK iteration consists of two CUDA kernels. The separate \texttt{IndexRemapUnifiedKernel} is not invoked for this workload: the fast-path remap for short sequences is handled inline inside \texttt{RadixSelectTopKUnifiedKernel}. Table~\ref{tab:dsa-topk-ncu-kernel-summary} summarizes the aggregate NCU metrics for the two kernels.

\begin{table}[htbp]
\centering
\small
\setlength{\tabcolsep}{3pt}
\caption{Per-kernel NCU summary for the slowest DSA-TopK workload.}
\label{tab:dsa-topk-ncu-kernel-summary}
\begin{tabular}{lcccc}
\hline
\textbf{Kernel} & \textbf{Time} & \textbf{SM} & \textbf{Occ.} & \textbf{Reg.} \\
 & \textbf{($\mu$s)} & \textbf{(\%)} & \textbf{(\%)} & \\
\hline
\texttt{FusedGemm} & 11.68 & 9.66 & 33.92 & 64 \\
\texttt{RadixSelect} & 13.18 & 0.16 & 23.54 & 40 \\
\hline
\multicolumn{5}{l}{\textbf{Total kernel time:} 24.86 $\mu$s} \\
\hline
\end{tabular}
\end{table}

To attribute time inside each fused kernel, we use NCU source-mapped PC sampling against \texttt{kernel.cu}. Table~\ref{tab:dsa-topk-phase-breakdown} reports the resulting phase-level breakdown. For the radix kernel, whose 4-CTA grid yields sparse PC samples, the phase split uses executed warp instructions as a proxy.

\begin{table}[htbp]
\centering
\small
\setlength{\tabcolsep}{3pt}
\caption{Source-attributed phase breakdown for DSA-TopK.}
\label{tab:dsa-topk-phase-breakdown}
\begin{tabular}{lc}
\hline
\textbf{Phase} & \textbf{Time ($\mu$s)} \\
\hline
\multicolumn{2}{l}{\textbf{\texttt{FusedGemm}}} \\
Entry / shared-memory setup & 2.88 \\
Scalar FP32 dot product & 2.88 \\
Q FP8-to-FP32 dequant & 1.86 \\
FP8 conversion helper & 1.52 \\
CUDA helper code & 1.52 \\
Page-table + K-scale load & 0.68 \\
K dequant + ReLU + reduce & 0.34 \\
\hline
\multicolumn{2}{l}{\textbf{\texttt{RadixSelect}}} \\
Fast-path index remap & 3.14 \\
Warp-ballot gather & 2.30 \\
Radix histogram build & 1.98 \\
Histogram zero-init & 1.27 \\
Suffix scan & 1.05 \\
Bucket-of-K selection & 0.83 \\
Score load + sortable encode & 0.59 \\
Histogram subgroup reduce & 0.36 \\
Slow-path setup & 0.31 \\
Other & 1.35 \\
\hline
\end{tabular}
\end{table}

The iteration is split roughly evenly between \texttt{FusedGemm} and \texttt{RadixSelect}. In \texttt{FusedGemm}, the largest costs are shared-memory/address setup and the scalar FP32 dot product. The GEMM share is modest because three of four batch elements satisfy \texttt{seq\_len <= TOPK} and return early, so only one batch element performs real score computation. In \texttt{RadixSelect}, the largest cost is the fast-path index remap for the three short batch elements, followed by warp-ballot gather and radix histogram construction.

\subsection{Dominant Kernel Drill-Down}

The two kernels expose different bottlenecks. \texttt{FusedGemmReluWSumUnifiedKernel} is limited by CTA-level early-return divergence, while \texttt{RadixSelectTopKUnifiedKernel} is limited by a very small cooperative grid.

\subsubsection{\texttt{FusedGemmReluWSumUnifiedKernel}}

The fused score kernel computes
\[
\mathrm{score}_{b,t} =
\sum_h \mathrm{ReLU}\!\left(Q_{b,h}K_t^\top\right) w_{b,h}.
\]
It uses a pure FP32 scalar dot product rather than tensor cores to preserve bit-exact agreement with the reference. Table~\ref{tab:dsa-topk-fused-gemm-sol} shows that the kernel reaches high L1/TEX activity but only 9.66\% SM throughput, indicating poor effective utilization rather than bandwidth saturation.

\begin{table}[htbp]
\centering
\small
\setlength{\tabcolsep}{4pt}
\caption{Speed-of-Light metrics for \texttt{FusedGemmReluWSumUnifiedKernel}.}
\label{tab:dsa-topk-fused-gemm-sol}
\begin{tabular}{lcc}
\hline
\textbf{Metric} & \textbf{Unit} & \textbf{Value} \\
\hline
Duration & $\mu$s & 11.68 \\
SM throughput & \% & 9.66 \\
Memory throughput & \% & 15.22 \\
DRAM throughput & \% & 0.37 \\
L1/TEX throughput & \% & 45.82 \\
L2 throughput & \% & 0.30 \\
SM frequency & GHz & 1.94 \\
DRAM frequency & GHz & 3.98 \\
\hline
\end{tabular}
\end{table}

Table~\ref{tab:dsa-topk-fused-gemm-launch-occ} reports the corresponding launch and occupancy statistics. Although the grid has 288 CTAs, only the CTAs associated with $b=2$ perform real GEMM work. The other batch elements early-return after the \texttt{seq\_len <= TOPK} check, leaving most CTAs idle.

\begin{table}[htbp]
\centering
\small
\setlength{\tabcolsep}{4pt}
\caption{Launch and occupancy metrics for \texttt{FusedGemmReluWSumUnifiedKernel}.}
\label{tab:dsa-topk-fused-gemm-launch-occ}
\begin{tabular}{lcc}
\hline
\textbf{Metric} & \textbf{Unit} & \textbf{Value} \\
\hline
Block size & threads & 512 \\
Grid size & CTAs & 288 \\
Waves per SM & -- & 0.97 \\
Registers per thread & reg/thread & 64 \\
Dynamic shared memory & KB/block & 48.13 \\
Theoretical occupancy & \% & 50.00 \\
Achieved occupancy & \% & 33.92 \\
Active warps per SM & warps & 21.71 \\
Warp cycles per issued inst. & cycles & 21.64 \\
Active threads per warp & -- & 31.30 \\
\hline
\end{tabular}
\end{table}

\subsubsection{\texttt{RadixSelectTopKUnifiedKernel}}

The radix kernel performs a four-pass 8-bit radix select to choose the top 2048 scores per batch. For sequences with \texttt{seq\_len <= TOPK}, it instead takes a fast path that writes the block-table-mapped indices and pads the rest with \texttt{-1}. Table~\ref{tab:dsa-topk-radix-topk-sol} shows that the radix kernel has extremely low SM and memory throughput, confirming that it is not compute- or bandwidth-saturated.

\begin{table}[htbp]
\centering
\small
\setlength{\tabcolsep}{4pt}
\caption{Speed-of-Light metrics for \texttt{RadixSelectTopKUnifiedKernel}.}
\label{tab:dsa-topk-radix-topk-sol}
\begin{tabular}{lcc}
\hline
\textbf{Metric} & \textbf{Unit} & \textbf{Value} \\
\hline
Duration & $\mu$s & 13.18 \\
SM throughput & \% & 0.16 \\
Memory throughput & \% & 0.87 \\
DRAM throughput & \% & 0.03 \\
L1/TEX throughput & \% & 11.42 \\
L2 throughput & \% & 0.12 \\
SM frequency & GHz & 1.95 \\
DRAM frequency & GHz & 3.98 \\
\hline
\end{tabular}
\end{table}

Table~\ref{tab:dsa-topk-radix-topk-launch-occ} explains this low utilization: the kernel launches only four CTAs, one per batch element, corresponding to 0.01 waves on 148 SMs. Three CTAs take the fast path, while only one CTA executes the full radix selection over 2241 keys.

\begin{table}[htbp]
\centering
\small
\setlength{\tabcolsep}{4pt}
\caption{Launch and occupancy metrics for \texttt{RadixSelectTopKUnifiedKernel}.}
\label{tab:dsa-topk-radix-topk-launch-occ}
\begin{tabular}{lcc}
\hline
\textbf{Metric} & \textbf{Unit} & \textbf{Value} \\
\hline
Block size & threads & 512 \\
Grid size & CTAs & 4 \\
Waves per SM & -- & 0.01 \\
Registers per thread & reg/thread & 40 \\
Dynamic shared memory & KB/block & 38.15 \\
Theoretical occupancy & \% & 50.00 \\
Achieved occupancy & \% & 23.54 \\
Active warps per SM & warps & 15.07 \\
Warp cycles per issued inst. & cycles & 17.79 \\
Active threads per warp & -- & 30.85 \\
\hline
\end{tabular}
\end{table}

\subsection{Summary}

The NCU profile confirms the main-text performance breakdown. On the slowest DSA-TopK workload, the two-kernel \texttt{FusedGemm} $\rightarrow$ \texttt{RadixSelect} pipeline is split roughly evenly in time. \texttt{FusedGemm} is CTA-divergence bound: three of four batch elements early-return, leaving only one batch element to perform real score computation. \texttt{RadixSelect} is cooperation-bound: its 4-CTA grid leaves most SMs idle, and only one CTA performs the full slow-path radix selection.

The remaining optimization headroom is therefore structural. A promising direction is to fold fast-path and slow-path work into a persistent or more globally scheduled kernel that keeps more SMs busy across the full iteration. Another possible direction is to use a tensor-core GEMM path for the long-sequence batch element while retaining the bit-exact scalar path only where needed by the benchmark constraints.

\end{document}